\documentclass[final]{svjour2}
\usepackage{graphicx}
\usepackage{rotating}
\usepackage{amssymb}
\usepackage{mathptmx}
\usepackage[numbers]{natbib}
\makeatletter
\journalname{Journal of Low Temperature Physics}

\bibpunct{}{}{,}{s}{}{,}

\begin{document}

\newcommand{\hdblarrow}{H\makebox[0.9ex][l]{$\downdownarrows$}-}
\title{Vortex Formations from Domain Wall Annihilations
 in Two-component Bose-Einstein Condensates}

\author{Hiromitsu Takeuchi$^1$ \and Kenichi Kasamatsu$^2$ \and Muneto Nitta$^3$\and Makoto Tsubota$^1$}

\institute{1:Department of Physics, Osaka City University, Sumiyoshi-Ku, Osaka 558-8585, Japan\\
Tel.:011+81-6-6605-2501\\ Fax:011+81-6-6605-2522\\
\email{hiromitu@hiroshima-u.ac.jp}
\\2: Department of Physics, Kinki University, Higashi-Osaka, Osaka 577-8502, Japan\\
 3: Department of Physics, and Research and Education Center for Natural Sciences, Keio University, Hiyoshi 4-1-1, Yokohama, Kanagawa 223-8521, Japan}

\date{\today}

\maketitle


\begin{abstract}

 We theoretically study the vortex formation from the collision of the domain walls in phase-separated two-component Bose-Einstein condensates.
 The collision process mimics the tachyon condensation for the annihilation of D-brane and anti-D-brane in string theory.
 A pair annihilation leaves the quantized vortices with superflow along their core, namely {\it superflowing cosmic strings}.
 It is revealed that the line density and the core size of the vortices depend on the initial distance between the walls.

\end{abstract}

\section{Introduction}
Topological defects or solitons appear not only in condensed matter physics but also in cosmology and high-energy physics.
It is interesting to connect the physical phenomena of topological defects in condensed matter to those in cosmology or high-energy physics \cite{BunkovGodfrin,Volovikbook},
 since the latter are difficult to be realized in experiments.
Atomic-gas Bose-Einstein condensates (BEC) are ideal systems for testing
 the physics of topological defects in theoretical and experimental aspects.
 One of the theoretical advantage is that the topological defects can
 be accurately described using the mean-field theory.
 Moreover, the development of the experimental technology enables us to control and directly visualize topological defects in the atomic clouds.

Recently, we proposed that analogues of D-branes \cite{Polchinski:1995mt},
 solitons in string theory,
can be realized in phase-separated two-component BECs \cite{KTNT},
 where the two systems are connected through the nonlinear sigma model \cite{Gauntlett,KTUreview}.
 In this model,
 the roles of D-branes and fundamental strings ending on them
 are played by domain walls and quantized vortices ending on the wall, respectively.
 There exist many advantages to study these solitons in this system,
 since it is possible to investigate experimentally and theoretically the dynamics of these solitons,
 such as oscillation modes \cite{Bretin, Simula} and instability \cite{Takeuchi_DGI, Takeuchi_KHI} of these solitons.
For example, Anderson {\it et al.} \cite{AHRFCCCsnake_exp} observed creation of vortex rings via dynamical (snake) instability of a dark soliton,
 where a nodal plane in one component was filled with the other component forming a pair of domain walls.
 In our context, this experiment can be regarded
 as a simulation of the string nucleation via the D-brane-anti-D-brane annihilation due to the tachyon condensation \cite{Sen_Tachyon}.

\begin{figure} [hbtp] \centering
  \includegraphics[width=.99 \linewidth]{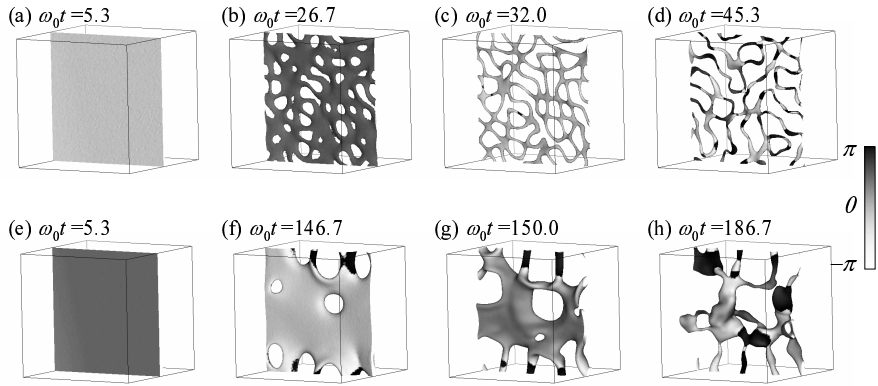}
  \caption{
 Dynamics of the vortex formations after the collision of the two domain walls.
 Figures show a normalized isosurface of the density difference $(n_1-n_2)/n^0=0.1$.
The upper and lower figures are the snap shots of the time evolutions from the stationary state with (a-d) $\mu_2=0.8 \mu_1$ and (e-h) $\mu_2=0.92 \mu_1$, respectively.
 The darkness of the surface plot refers to the phase $S_2$ of the condensate wave function of the component $2$, which is trapped along the vortex cores of the component $1$ [(d) and (h)].
 The box size is $(51.2\times \xi)^3$ with the healing length $\xi=\hbar/\sqrt{mgn^0_1}$.
 The unit of time is $1/\omega_0=\hbar/\mu_1$. 
}
\label{fig:annihilation}
\end{figure}

 In this paper, we study the vortex formation
 by a pair annihilation of domain walls of phase-separated two-component BECs in uniform systems.
The configuration of the subsequent vortex rings was strongly affected by the geometry of the trapping potential in the experiment \cite{AHRFCCCsnake_exp}.
Since we are not interested in such a finite size effect,
 we confine ourselves to an uniform system and show that vortex formation becomes more complex (see Fig. \ref{fig:annihilation}).
 The vortex formations are strongly affected by the existence of the filling component,
 which was neglected in the experiment  \cite{AHRFCCCsnake_exp}.
 The filling component strongly affects on the line density and the core size of vortices nucleated after the annihilation.
 
\section{D-brane and anti-D-brane in two-component BECs}
\label{sect:SimpleModel}
 The two-component BECs are described by the condensate wave functions (order parameters) $\Psi_j=\sqrt{n_j}e^{iS_j}$ in the mean-field approximation at $T=0$,
 where the index $j$ refers to the component $j$ ($j=1,2$).
The wave functions obey the coupled Gross-Pitaevskii (GP) equations \cite{Pethickbook}
\begin{eqnarray}
i \hbar \frac{\partial \Psi_j}{\partial t} = \left[-\frac{\hbar^2}{2m_j}{\bf \nabla}^2+\sum_k g_{jk}|\Psi_k|^2\right]\Psi_j,
\label{eq:GP}
\end{eqnarray}
where we use the particle masses $m_j$, 
the intra-atomic interaction parameters $g_{jj}$,
 and the inter-atomic interaction parameter $g_{12}$.
 When the condition $g_{12}>\sqrt{g_{11}g_{22}}$ is satisfied \cite{Timmermans},
 the two condensates undergo the phase separation making the interface layer between domains.
 In the domain of the component $j$ far from the interfaces, we have $n_j \approx n^0_j\equiv\mu_j/g_{jj}$ and $n_k\approx 0~(k\neq j)$.
 We can regard a plane $n_1=n_2$ in the interface layer as an analogue of D-brane \cite{KTNT}.
 We use the parameters $m=m_1=m_2$ and $g=g_{11}=g_{22}=0.5g_{12}$ similar to Ref. \cite{KTNT},
 where the interface thickness is $\sim \xi=\hbar/\sqrt{mgn_j^0}$.

Let us consider two flat interfaces perpendicular to the $z$-axis,
 where a domain of the component $2$ is between a D-brane at $z=-R/2 \leq 0$ and an anti-D-brane at $z=R/2 \geq 0$.
 If the distance $R$ between the D-brane and the anti-D-brane is large enough,
 the correlation between  the two domains of the component $1$ in $z<-R/2$ and $z>R/2$ is negligible,
 and the phase difference $dS_1\equiv S_+-S_-$ between the two domains is
 arbitrary determined,
 where we defined the phases as $S_1=S_-$ in $z<-R/2$ and $S_1=S_+$ in $z>R/2$.

When the two interfaces are close to each other,
 the phase difference $dS_1\equiv S_+ - S_-$ becomes important.
 Since the phase $S_1(z)$ must continuously change in $-R/2<z<R/2$,
 the phase difference $dS_1\neq 0$ yields the gradient energy from the spatial variation of $S_1(z)$ after the collision of the two interfaces.
 If the phases $-\pi \leq S_1(z) \leq \pi$  are mapped to a unit circle,
 there are two ways from $S_-$ to $S_+$, clockwise or counterclockwise.
 The shorter routes are preferred to decrease the gradient energy.
Especially, for $|dS| = \pi$, there are two possible routes from $S_-$ to $S_+$.
 The possibility of different routes can leave topological defects along the plane,
 at which the interfaces collide.
 This situation is similar to that of the dynamical (snake) instability of nodal planes (dark solitons) for single-component BECs in two or three dimensional systems.
 However, the instability would be suppressed for sufficiently large $R$ due to the presence of the component $2$, with which the nodal plane is filled.

 \section{Emergence of excitations with complex frequencies} 
 To reveal such an effects on the vortex formation,
we first investigate the linear stability of the dark soliton of the component $1$ whose density dip is filled with the component $2$,
 which is called dark-bright soliton \cite{BuschAnglin}.
 Then the condensate wave functions in a stationary state may be written as $\Psi^0_j=\psi_j(z)\exp (-i\mu_jt/\hbar)$ with real functions $\psi_j$.
 The excitations are expressed by the perturbations $\delta \Psi_j=\Psi_j-\Psi^0_j$ from the stationary state.
 Because of the symmetry of the stationary state $\Psi^0_j$,
 we have $\delta \Psi_j={\cal U}^+_j-{{\cal U}^-_j}^*$ with
 ${\cal U}^{\pm}_j({\bf r},t)=u^{\pm}_j(z)\exp\left(ikx-i\omega t \mp i\mu_j t/\hbar \right)$,
 where $\mu_j$ is the chemical potential of the component $j$,
 where we may neglect the coordinate $y$ for simplicity in the linear analysis.
 By linearizing the GP equations (\ref{eq:GP}) with respect to $\delta \Psi_j$,
 we obtain the reduced Bogoliubov-de Gennes (BdG) equations
 \begin{eqnarray}
\hbar \omega {\bf u}
=
\hat{\cal H}{\bf u},~~~
\hat{\cal H}
=
\left( 
\begin{array}{cccc}
\hat{h}_1^+ & -g_{11}\psi_1^2 &g_{12}\psi_2\psi_1 & -g_{12}\psi_2\psi_1 \\
g_{11}\psi_1^2 & -\hat{h}_1^- &g_{12}\psi_2\psi_1 & -g_{12}\psi_2\psi_1 \\
g_{12}\psi_1\psi_2 & -g_{12}\psi_1\psi_2 & \hat{h}_2^- & -g_{22}\psi_2^2 \\
g_{12}\psi_1\psi_2 & -g_{12}\psi_1\psi_2 & g_{22}\psi_2^2 & -\hat{h}_2^+ \\
\end{array} 
\right)
\label{eq:BdG}
\end{eqnarray}
 where ${\bf u}=(u_1^+, u_1^-, u_2^+, u_2^-)^T$ and
$\hat{h}^{\pm}_j=
\frac{\hbar^2k^2}{2m}-\frac{\hbar^2}{2m}\frac{d^2}{dz^2}+g\psi_j^2-\mu_j+\sum_i g_{ji}\psi_i^2$.
 When the frequency has a non-zero imaginary part ${\rm Im}[\omega(k)]\neq 0$, the mode is exponentially amplified.
  We can obtain $\Psi^0_j$, $\delta \Psi_j$ and $\omega$ by numerically solving the GP Eqs. (\ref{eq:GP}) and the BdG Eqs. (\ref{eq:BdG}).
\begin{figure} [hbtp] \centering
  \includegraphics[width=.99 \linewidth]{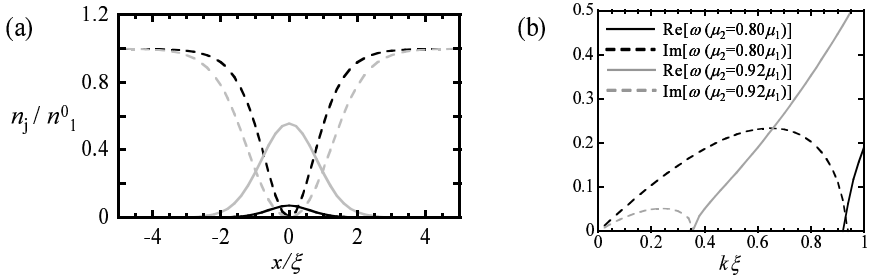}
  \caption{
(a) The density profiles of the phase-separated two-component condensates with two interfaces.
[(black broken line) $n_1$ and (black solid line) $n_2$ for $\mu_2=0.80\mu_1$,
 and (gray broken line) $n_1$ and  (gray solid line)$n_2$ for $\mu_2=0.92\mu_1$].
 The component $1$ has a node and $n_1=0$ at $z=0$. 
 The units of the vertical and horizontal axes are $n^0_j=\mu_j/g$ and $\xi=\hbar/\sqrt{mg n^0_1}$, respectively. 
(b) The dispersion relations of the interface modes in the states of (a).
 The units of the vertical and horizontal axes are $\omega_0=\mu_1/\hbar$ and $1/\xi$, respectively. 
}
\label{fig:linear}
\end{figure}

 Figures \ref{fig:linear} shows the profiles $\Psi^0_j$ and the dispersion $\omega(k)$ for typical two values of $\mu_2/\mu_1<1$.
 The particle number of the component $2$, thus, the distance $R$ between the branes increases with $\mu_2/\mu_1$.
 For $\mu_2/\mu_1<1$, the density $n_2$ of the filling component is smaller than the bulk value $n^0_2$ at $z=0$.
 The imaginary part ${\rm Im}[\omega(k)]\neq 0$ appears for a range of $k$ and
 has its maximum at the wave number $k_{\rm max}$.
 The maximum value monotonically decreases
 and the dispersion $\omega(k)$ becomes eventually real for $\mu_2/\mu_1 \to 1$.
In other word,
 the excitations with complex frequencies gradually emerge when the D-brane and the anti-Dbrane approach to each other for $\mu_2/\mu_1<1$.

 The emergence of modes with complex frequencies means that the system is dynamically unstable.
 The unstable modes are exponentially amplified in the linear stage of the instability.
 The amplification of the unstable modes with finite wave numbers $k\neq 0$ induces superflows,
 where the atomic interaction energy is transduced to the gradient energy.
 As a result, the instability leads to the vortex formations as shown Fig. \ref{fig:annihilation}.
 Such a process of the instability due to the unstable mode with complex frequency
 mimics the tachyon condensation caused by the tachyonic mode with complex mass $m^2<0$ in string theory \cite{Sen_Tachyon}.
 The tachyon condensation can leave lower dimensional topological defects after the annihilation of D-brane and anti-D-brane.
 In our case of the phase-separated two-component BECs,
 the annihilation of the $2$-dimensional defects (domain walls) leaves $1$-dimensional defects (quantized vortices).
 Since the time scale of the amplification of a `tachyonic' mode is given by $\tau_d \sim 1/{\rm Im}[\omega(k)]$,
 the modes with the wave number $\sim k_{\rm max}$ are dominantly amplified,
 which should be related to the vortex formations.

\section{Vortex formation from the annihilation of  the D-brane and the anti-D-brane}
\begin{figure} [hbtp] \centering
  \includegraphics[width=.99 \linewidth]{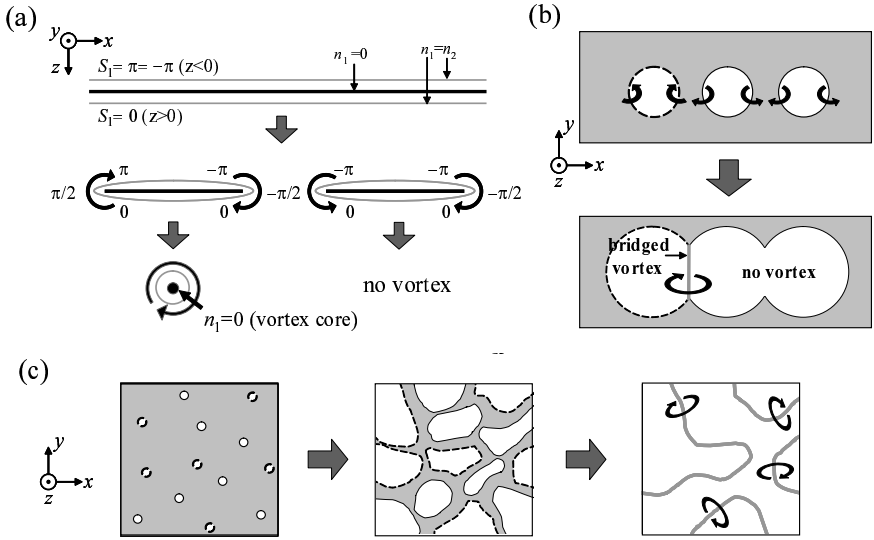}
  \caption{
Schematic diagrams of the vortex formation in the collision process.
(a) The planer object bounded between the D-brane and the anti-D-brane transforms into a quantized vortices.
 The top figure shows the flat planer object in a stationary state.
 Black and gray lines show the nodal plane with $n_1 =0$ (inner wall) and D-brane (or anti-D-brane) with $n_1=n_2$, respectively.
 (b) The reconnections of the rings of the `half-quantized' vortex living on the planer object.
 The sign of the vortex ring (tunnel) is represented with solid or broken line.
The gray background refers to the planer object and the white regions represent the tunnels through the object.
 The reconnection between the rings with different signs yields a bridged vortex.
(c) Vortex formation after the growth of the rings with random distribution.
 The random distribution causes a distorted vortex line formation.
}
\label{fig:image}
\end{figure}
 In the stationary state without the unstable modes, the nodal ($n_1=0$) plane exists between the branes ($n_1=n_2$ planes).
 Then the region with $n_1 \leq n_2$ bounded between the branes forms a planer object [Fig. \ref{fig:image}(a) top].
 When the unstable modes are excited and amplified locally,
 the planer object is broken and there appear tunnels through the planer object as shown in Fig. \ref{fig:image}(a) (middle).
 The tunnels are characterized by the direction of the phase gradient perpendicular to the planner object.
 Since the phase $S_1$ changes from $0$ to $\pi$ or $-\pi$ through the tunnel,
 the `edge' of the tunnel form a pseudo-vortex which mimics a half-quantized vortex.
 Although a pair of the the pseudo-vortices with different circulations annihilates and destroys the $n_1 \leq n_2$ region,
 a pair of the vortices with the same circulations leaves the $n_1 \leq n_2$ region as a single-quantized vortex whose core is filled with the component $2$.
 These processes look like reconnections between the rings of the half-quantized vortex,
 which live on the planer object [Fig. \ref{fig:image}(b)].
 Then the reconnections make a larger rings or leave a bridge of single-quantized vortex.

 Since the initial perturbation can be taken to be arbitrary in the uniform systems,
 the positions and the signs of the tunnels would be distributed randomly [Fig. \ref{fig:image}(c) left].
 Then, the growing tunnels form the mesh structures [Fig. \ref{fig:image} (c) (center)],
 which causes a distorted vortex line formation [Fig. \ref{fig:image} (c) (right)].

Now, we can estimate the vortex line density $n_v$ after the collision.
Since the unstable modes with $k\sim k_{\rm max}$ are dominantly amplified,
 the mean distance between the neighboring vortex ring would be of order $\sim 1/k_{\rm max}$ in the early stage of the instability.
 As a result, we have the vortex line density $n_v\sim k_{\rm max}$ per unit area along the $z=0$ plane.
 Since $k_{\rm max}$ is decreased with $\mu_2/\mu_1$,
  the resultant vortex line density increases with $\mu_2/\mu_1$.

 These interpretations explain well the numerical results in Fig. \ref{fig:annihilation}.
 The numerical simulation was done by solving the GP Eqs. (\ref{eq:GP}) from the stationary states $\Psi^0_j$ for $\mu_2/\mu_1=0.8,~0.92$ with a random seed for the initial perturbation in a periodic system along the $x$-$y$ plane.
 Since the particle number is conserved in these simulations,
 the particle density of the component $2$ per unit length along the vortex line becomes larger for weak instability with small $k_{\rm max}$.
 The fluctuation of the phase $S_2$ of the component $2$ causes the superflow along the vortex line.
 Such a vortex may be called as a core-flow vortex or {\it a superflowing cosmic string}, analogue of the superconducting cosmic strings \cite{Witten}.
 For $\mu_2/\mu_1=0.92$ in Figs. \ref{fig:annihilation} (bottom),
 the particle number of the component $2$ is so large that we cannot define well the vortex core in contrast to the case of $\mu_2/\mu_1=0.80$ in Figs. \ref{fig:annihilation} (top).

 In the experiment by Anderson {\it et al.} \cite{AHRFCCCsnake_exp}, the snake instability was induced by removing the filling component selectively.
 This experimental technique may control the relative velocity between the D-brane and the anti-D-brane.
 If they approach to each other slowly,
 the dynamical instability emerges gradually and thus the vortex formations are suppressed.
 The vortex line density $n_v$ after the annihilation increases with the relative velocity.
 Roughly speaking, the density $n_v$ takes the maximum value $\sim k_{\rm max}(\mu_2=0) \sim \xi^{-1}$ when the relative velocity is larger than $\sim \xi/\tau_{\min}$,
 where $\tau_{\min}= 1/{\rm Im}[\omega(k_{\max})]$ is the growth time of the instability for $\mu_2/\mu_1=0$ and $\xi$ is the thickness of the interface layers.

\section{Conclusion}
We have studied the vortex formation after the pair annihilation of the domain wall and the anti-domain wall in phase-separated two-component BECs in uniform systems.
 The vortex line density after the annihilation is qualitatively estimated,
 which may reach to its maximum value for a sufficiently large relative velocity between the walls.
 The nucleated vortices after the annihilations have superflows along their core regions. 
 These phenomena can be experimentally realized in phase-separated condensates
 whose size along the walls is larger than the length scale $\sim 1/k_{\rm max}$.
 The detailed analysis of this study will be reported elsewhere soon.

 It is interesting to mention about the relation to the defect creations by colliding interfaces in the experiments of superfluid ${}^3$He \cite{BFGHKMPRT_NatPhys2008},
 where A-phase domain is between two B-phase domains forming the A--B and B--A interfaces as analogues of branes. 
 The underlying mechanisms of vortex formations proposed in this work might be applied to the vortex formation in the ${}^3$He experiments.
 In our case of two-component BEC system, vortex formations are essentially triggered by making tunnels in an unstable soliton of the $1$st component {\it hidden} in the domain of the $2$nd component.
 Since such an unstable soliton is possible in ${}^3$He-B \cite{Volovikbook},  
we may expect that the similar vortex formations occur in the pair annihilations of the A--B and B--A interfaces in the ${}^3$He experiments if the soliton is {\it hidden} in the A-phase domain.
 The theoretical and numerical analysis in this direction are the interesting future works.

\begin{acknowledgements}
This work was supported by KAKENHI from JSPS
(Grant No. 21740267, 199748, 20740141, and 21340104)
and from MEXT (Grant No. 17071008).
\end{acknowledgements}



\begin{thebibliography}{99}

\bibitem{BunkovGodfrin}
{\it Topological Defects and the Non-Equilibrium Dynamics of Symmetry
Breaking Phase Transitions}, ed. Y. M. Bunkov and H. Godfrin (Kluwer Academic, Dordrecht, 2000) NATO Science Series C 549.

\bibitem{Volovikbook}
G. E. Volovik, {\it The Universe in a Helium Droplet} (Clarendon Press, Oxford, 2003).

\bibitem{Polchinski:1995mt}
J. Polchinski, Phys. Rev. Lett.  {\bf 75}, 4724 (1995).


\bibitem{KTNT}
K. Kasamatsu, H. Takeuchi, M. Nitta, and M. Tsubota, arXiv:1002.4265.


\bibitem{Gauntlett}
J. P. Gauntlett, R. Portugues, D. Tong, and P. K. Townsend,
 Phys. Rev. D {\bf 63}, 085002 (2001). 

\bibitem{KTUreview} 
K. Kasamatsu, M. Tsubota, and M. Ueda, Int. J. Mod. Phys. {\bf 19}, 1835 (2005).

\bibitem{Bretin}
V. Bretin, P. Rosenbusch, F. Chevy, G. V. Shlyapnikov, and J. Dalibard,
 Phys. Rev. Lett. {\bf 90}, 100403 (2003).

\bibitem{Simula}
T. P. Simula, T. Mizushima, and K. Machida,
Phys. Rev. Lett. {\bf 101}, 020402 (2008).


\bibitem{Takeuchi_DGI}  
H. Takeuchi, K. Kasamatsu, and M. Tsubota,
Phys. Rev. A {\bf 79}, 033619 (2009).

\bibitem{Takeuchi_KHI}  
H. Takeuchi, N. Suzuki, K. Kasamatsu, H. Saito, and M. Tsubota,
Phys. Rev. B {\bf 81}, 094517 (2010).

\bibitem{AHRFCCCsnake_exp} 
B. P. Anderson, P. C. Haljan, C. A. Regal, D. L. Feder, L. A. Collins, C.W. Clark, and E. A. Cornell,
Phys. Rev. Lett. {\bf 86}, 2926 (2001).

\bibitem{Sen_Tachyon}
Ashoke Sen,
 Int. J. Mod. Phys. A {\bf 20}, 5513 (2005).


\bibitem{Pethickbook}
C.J. Pethick and H. Smith, {\it Bose-Einstein Condensation in Dilute Gases, 2nd ed.}, Cambridge University Press, Cambridge (2002).

\bibitem{Timmermans} 
E. Timmermans, Phys. Rev. Lett. {\bf 81}, 5718 (1998).

\bibitem{BuschAnglin} 
Th. Busch and J. R. Anglin, Phys. Rev. Lett. {\bf 87}, 010401 (2001).

\bibitem{Witten}
 E. Witten, Nucl. Phys. B {\bf 249}, 557 (1985).
 
\bibitem{BFGHKMPRT_NatPhys2008}
 D. I. Bradley, S. N. Fisher, A. M. Gu$\acute{\rm e}$nault, R. P. Haley, J. Kopu, H. Martin, G. R. Pickett, J. E. Roberts, and V. Tsepelin, Nature Phys. {\bf 4}, 46 (2008).


\end{thebibliography}
\end{document}